      [1999/12/01 v1.4c Il Nuovo Cimento]
\documentclass[varenna]{cimento}

\usepackage{graphicx}  % got figures? uncomment this
\title{Production of a degenerate Fermi gas of metastable helium-3 atoms}
\author{W.~Vassen, T.~Jeltes, J.~M. McNamara \atque A.~S. Tychkov}
\institute{Laser Centre Vrije Universiteit, De Boelelaan 1081, 1081 HV Amsterdam, The Netherlands}
\begin{document}

\maketitle

\section{Introduction}
Most of this book discusses Fermi gases composed of alkali atoms. Since the first realization of
a degenerate $^{40}$K Fermi gas in 1999, $^6$Li has been the only other fermionic
atom to be cooled below the Fermi temperature. In this contribution we will
discuss our recent results on cooling fermionic $^3$He below the Fermi temperature.
This has not been performed with ground state $^3$He, but with $^3$He in the metastable
$2\,^3\rm{S}_1$ state, which has an internal energy of 20 eV. Helium atoms in this metastable state 
are denoted by He* and have a lifetime
of $\sim 8000\,\rm{s}$, which is infinite for all practical purposes. He* atoms can be 
efficiently cooled with laser light at 1083\,nm and samples of $\sim10^9$ atoms
of either isotope can be confined in a magneto-optical trap (MOT) at temperatures
around 1\,mK. 
The density in a He* MOT is limited to $\sim4\times10^{9}$\,cm$^{-3}$ due to
large losses associated with Penning ionization (PI) and 
associative ionization (AI):
\begin{equation}
\begin{array}{lr}
\textup{He*}+\textup{He*}\rightarrow \textup{He}+\textup{He}^++e^- &
\mbox{(PI)}\\
%\\
%\nonumber 
\textup{He*}+\textup{He*}\rightarrow \textup{He}^{+}_2+e^- &
\mbox{(AI)}
\end{array}
\label{Eq:PI}
\end{equation}
These loss processes (summarized as Penning Ionization in the rest of this contribution) 
are discussed in many studies of ultracold metastable helium and are both a
strength and a weakness of this atom in cold atom experiments. To begin with the weak side, the 
rate constant for ionizing collisions in a typical MOT is 
$\sim5\times10^{-9}$\,cm$^3$/s
and it is this rate that limits the achievable densities in a MOT and forces He* MOTs to
have a large diameter (up to 1\,cm) in order to accumulate a large number of atoms. 
When the light is switched off this loss rate is reduced to
$\sim10^{-10}$\,cm$^3$/s \cite{Stas06}, still far too large for further cooling processes lasting seconds.
It is only when the atoms are fully spin-polarized that the rate constant drops by four
orders of magnitude \cite{Shlyapnikov94} and becomes acceptable. Before our work 
on cooling and trapping $^3$He* \cite{Stas06, Stas04, McNamara06} this 
suppression of Penning ionization had only been demonstrated
for the bosonic isotope $^4$He, which, due to this reduced loss rate constant, could be brought 
to Bose-Einstein condensation (BEC) \cite{Robert01, Pereira01, Tychkov06}. The reduction in loss rate can be simply understood
by considering the angular momentum quantum numbers (for $^4$He*) in more detail:
\begin{equation} 
\textup{He*}({\it J}=1)+\textup{He*}({\it J}=1) \rightarrow 
\textup{He}({\it J}=0) + \textup{He}^+({\it J}=1/2) + e^-({\it J}=1/2) 
\label{Eq:PIsuppression}
\end{equation}
The $2\,^3\rm{S}_1$ state has a total angular momentum (spin) $J=1$. In a collision between two 
He* atoms a
total angular momentum of 0, 1 and 2 is therefore possible, while 
for the reaction products the maximum combined angular momentum is 1. 
Hence, Penning ionization is forbidden (as long as spin is conserved) in either of the
two cases in which the atoms are spin-polarized in either of the fully stretched
magnetic substates ($m=+1$ or $m=-1$) for which $J=2$.
Experimental studies have to date concentrated on the $m=+1$ state which is
magneticallly trappable.
Only due to the spin-dipole interaction Penning ionization may occur, and 
the rate constant for that process is $\sim10^{-14}$\,cm$^3$/s \cite{Shlyapnikov94}.
Consequently, a mixture containing $m\neq+1$ atoms will show strong Penning  
ionization.

\begin{figure}
\begin{center}
\scalebox{0.8}{\includegraphics{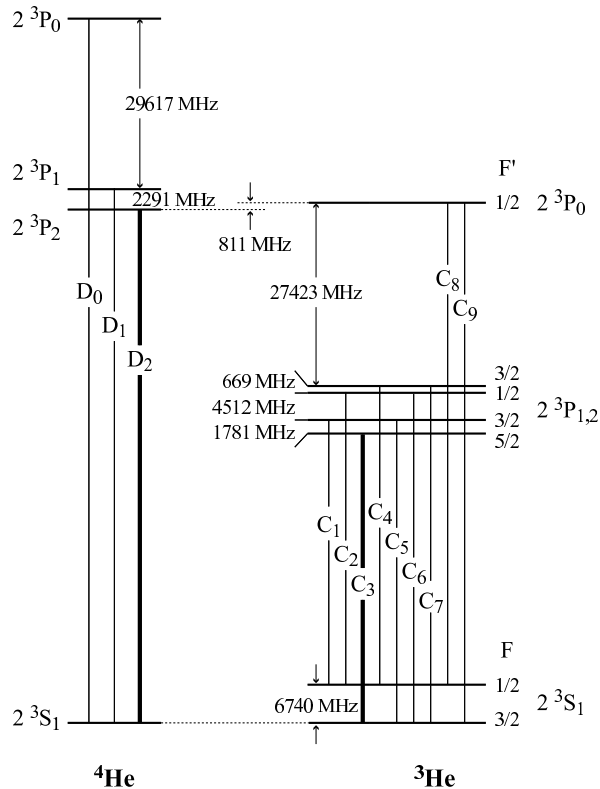}}     % includes figure foo.eps
\end{center}
\caption{$^4$He and $^3$He energy levels relevant for laser cooling 
and probing at 1083\,nm.
\label{A1}}
\end{figure}

In contrast to the $^4$He case, the fermionic isotope $^3$He
has nuclear spin $I=1/2$, which makes the level structure more complex. Due to this
nuclear spin the metastable $2\,^3\rm{S}_1$ state is split by the hyperfine interaction
into an $F=3/2$ and $F=1/2$ state (splitting 6740\,MHz) with 
the $F=3/2$ state having the lowest energy (see Fig.~\ref{A1}). 
When we consider magnetic trapping,
three magnetic substates are in principle trappable, 
i.e., $|F,M_F\rangle=|1/2,+1/2\rangle$, $|3/2,+1/2\rangle$ and 
$|3/2,+3/2\rangle$.
In a magnetic trap we may trap atoms in any of these states, although PI will strongly 
reduce the lifetime of any mixture of these spin states. The Pauli exclusion principle,
however, suppresses collisions between identical fermions and PI should be suppressed in a cloud of 
$^3$He* atoms in a single spin state, though this would come at the cost of rethermalizing 
collisions and our ability to evaporatively cool such a sample.
Sympathetic cooling overcomes this limitation by introducing a second component to
the gas; thermalization between the two components then allows the mixture as a whole to be cooled.
As sympathetic cooling usually works best in a mixture
with bosonic atoms, we need to consider bosons that do not
Penning ionize in collisions with $^3$He*. An obvious choice for us was $^4$He* as
the experimental setup for evaporative cooling was available. However,
we can not use all three trappable $^3$He* states now. Although there have been no
theoretical predictions, we may expect that only in collisions between
the $|3/2,+3/2\rangle$ state of $^3$He* and the 
$|J,m_J\rangle=|1,+1\rangle$ state of $^4$He*
the same reduction of Penning ionization as in a pure gas of 
$|1,+1\rangle$ atoms of $^4$He*
will occur because of conservation of angular momentum as above. 
Of course this may have considerable consequences for further studies of
ultracold $^3$He* clouds and $^3$He*/$^4$He* mixtures. For the mixtures only a single spin
combination is expected to be stable, i.e., $^3$He* in the 
$|3/2,+3/2\rangle$ state and $^4$He in the
$|1,+1\rangle$ state. For an ultracold $^3$He* gas there are more options, however no mixtures
of different magnetic substates. This hampers studies where one controls the scattering
length via Feshbach resonances. Although no studies of possible Feshbach resonances
have so far been performed, it may be expected that even if there were magnetic fields at which
they existed, the required mixture of different hyperfine states 
($|1/2,+1/2\rangle$ and $|3/2,+1/2\rangle$
for instance) would be very unstable. One should not draw the conclusion that
only the $^3$He* $|3/2,+3/2\rangle$ - $^4$He $|1,+1\rangle$ system would be suitable for producing
and studying an ultracold gas of $^3$He* atoms. Any bosonic alkali atom in a {\it fully stretched}
hyperfine state may show the suppression of Penning ionization in collisions with
$^3$He* atoms in the $|3/2,+3/2\rangle$ state. It remains to be seen, of course, if this suppression
will be as large as observed in collisions with $^4$He $|1,+1\rangle$ atoms.  

From an experimentalists point of view the Penning ionization process provides new and exciting
opportunities to investigate cold clouds of He* atoms.
First, as the dominant loss process produces ions, we have a direct
and 'nondestructive' detection method. Ions can be efficiently detected with an electron 
multiplier: when an ion hits a microchannel plate (MCP) detector with sufficient
kinetic energy, an electron is released which can be amplified and detected. When the count rate 
is not too high one can even count individual ions. Second, in the same way as ions can be 
detected also neutral He* atoms can be detected: the 20\,eV internal energy far exceeds the electron
work function of almost all materials and the released electron can be counted by a 
similar MCP detector. The application of He$^+$ and He* detectors allows alternative
means of observation of an ultracold cloud of metastable helium, besides the standard method
of absorption imaging. The ion detector is effectively a density probe (see Eq.~\ref{Eq:PI}) 
while the He* detector
measures the arrival time (and position in the case of a position-sensitive MCP) of atoms after
release or escape from the trap.
\begin{figure}[b]
\begin{center}
\scalebox{0.8}{\includegraphics{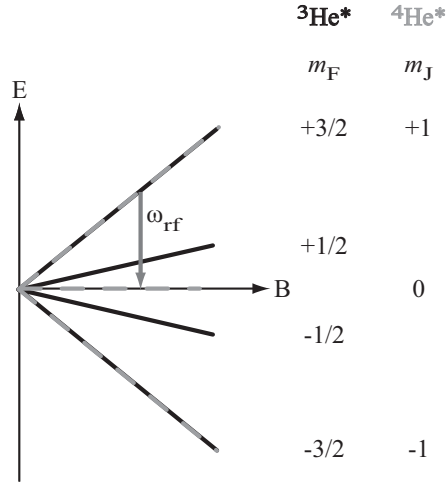}}     % includes figure foo.eps
\end{center}
\caption{Magnetic field dependence of the $^3$He* $F=3/2$ and $^4$He* $J=1$ magnetic substates.
Evaporative cooling, at RF frequency $\omega_{rf}$, is performed on 
the $M_J=+1\rightarrow M_J=0$ transition.}
\label{Zeeman}
\end{figure}
 
\section{Relevant atomic physics}
In contrast to all alkali cold atom research, experiments on metastable helium are performed with
atoms in the highly excited metastable $2\,^3\rm{S}_1$ state which 
lies 20\,eV above the $1\,^1\rm{S}_0$ ground state.
The ionization limit is at 25\,eV.
Collisions between He* atoms therefore contain 40\,eV internal energy which is more than enough to
ionize one of the two collision partners, while the other is de-excited (see Eq.~\ref{Eq:PI}). 
Atoms in the $2\,^3\rm{S}_1$ state can be laser cooled by applying 
radiation at 1083\,nm to excite
the $2\,^3\rm{P}$ state. In Fig.~\ref{A1} the energy level structure relevant for laser cooling is summarized.
For $^4$He the D$_2$ transition, $2\,^3\rm{S}_1 \rightarrow 
2\,^3\rm{P}_2$, that is commonly
used for efficient cooling and trapping in a MOT, is indicated, whilst for $^3$He
cooling is performed on the C$_3$ ($F=3/2 \rightarrow 5/2$) transition
to prevent optical pumping to the $F=1/2$ hyperfine state.
The more than 33\,GHz frequency difference between the two
laser cooling transitions implies that the laser radiation for cooling both isotopes cannot
come from one laser. Moreover, there is a near coincidence between the $^4$He laser 
cooling transition and the C$_9$ transition
of $^3$He, which may cause unwanted optical pumping. However, when only cooling $^3$He* in the
$F=3/2$ state, no repumper is required~\cite{Stas04}.

The Zeeman splitting of the metastable states
involved in magnetic trapping and evaporative cooling is shown in Fig.~\ref{Zeeman}. 
The fully stretched states of both isotopes
show the same magnetic field dependence. For sympathetic cooling of
$^3$He* by $^4$He*, evaporative cooling of $^4$He* is performed on the $M_J=1\rightarrow0$
transition. The corresponding RF radiation will then not affect the $^3$He* atoms
(at the same temperature) as these are resonant at lower frequency. Due to the large hyperfine
and fine-structure splittings the magnetic field dependence is linear in magnetic
traps typically used in BEC experiments of $^4$He*.

\section{Molecular and collision physics}   
\begin{figure}[b]
\begin{center}
\scalebox{1.0}{\includegraphics{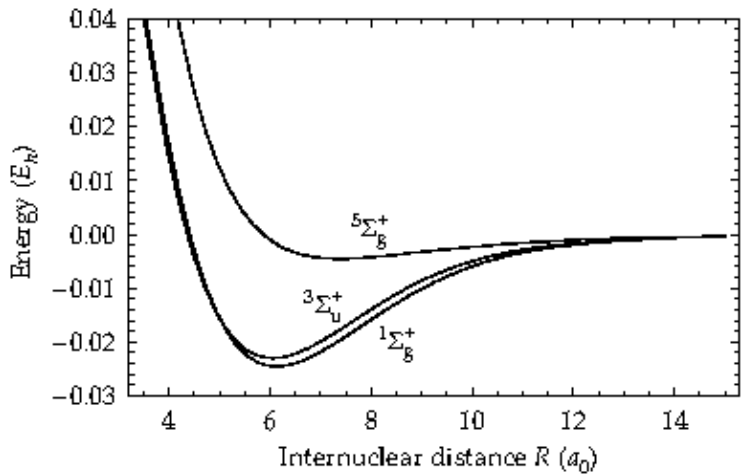}}     % includes figure foo.eps
\end{center}
\caption{Short-range potentials in atomic units.}
\label{shortrangepot}
\end{figure}

Collisions between two He* atoms are described
with interatomic interaction potentials that can be calculated {\it ab initio} in
the Born-Oppenheimer approximation, where the total electronic spin $S$ is a good quantum
number. The possible values $S=0$, 1, and 2 correspond to a singlet, triplet and quintet
potential (see Fig.~\ref{shortrangepot}).
In a MOT the atoms are unpolarized and collisions take place in all three potentials.
In a magnetic trap with atoms in the fully stretched state, collisions take place in only the
quintet potential and the binding energy of the least bound state determines the
scattering length. The quantum chemistry calculations of the quintet potential
have recently become so good that the scattering length
for collisions between spin-polarized $^4$He* atoms can be calculated with an accuracy
of 3\%. The calculated value is $a_{44}=+7.64(20)$\,nm \cite{Przybytek05}, which 
agrees very well with an experimental value $a_{44}=+7.512(5)$\,nm recently
obtained in Paris \cite{Moal06}. The value for $a_{44}$ is relatively large, about 50\% larger than 
corresponding numbers for Rb and Na. This means that evaporative cooling of $^4$He*
proceeds efficiently and short RF ramps can be used. The accuracy of the theoretical
calculations suggests that similar accuracy can be obtained when calculating the 
molecular potentials between two spin-polarized $^3$He* atoms, and, more importantly,
for one spin-polarized $^3$He* atom and one spin-polarized $^4$He* atom. These 
calculations have recently been performed as well. A value
$a_{34}=+28.8^{+3.9}_{-3.3}$\,nm \cite{Jeziorski05} results, which is positive and exceptionally
large. The large value implies that sympathetic cooling of $^3$He*
by $^4$He* in a magnetic trap should proceed very efficiently. Moreover, an ultracold
mixture of spin-polarized $^3$He* and $^4$He* atoms should be stable as
the interaction is repulsive. Of course
it remains to be seen what the two-body and three-body loss rate constants will be
for such a large heteronuclear scattering length. Calculations have not yet been performed, but an
order of magnitude estimate of the three-body loss rate constant $K^{BBF}_3$
for collisions between two $^4$He* bosons and 
one $^3$He* fermion, which is expected to be the largest, gives
$K^{BBF}_3\approx1\times10^{-24}$\,cm$^6$/s~\cite{McNamara06, Esry04}. This value is 1-3 orders of magnitude
larger than the three-body loss rate of ultracold $^4$He* alone. 

\section{MOT results}
We load our MOT with either $^4$He*, $^3$He*, or an arbitrary mixture of the two.
We start with a DC discharge to populate the metastable state. The discharge is
maintained inside our source chamber. The helium mixture is prepared outside the vacuum
chamber and coupled via a needle valve into a discharge tube which is positioned
1-2\,cm in front of a skimmer with a 1\,mm hole. The diverging He* beam is collimated
over a length of 20\,cm by applying transversal laser cooling in two dimensions 
to increase the beam intensity~\cite{Rooijakkers96}, and deflected over an angle of 1 degree
to purify the He* beam. The latter is important to prevent
ground state atoms, VUV photons and singlet metastables (from the source) reaching the ultrahigh
vacuum (UHV) chamber in which the final cooling and trapping experiments take place.
After the collimation/defection zone the He* beam enters a 2\,m Zeeman slower which
slows the atoms to a velocity of $\sim$70\,m/s, which is low enough to capture them
in a MOT. The MOT configuration is standard except that very large MOT beams have
to be used in order to trap as many atoms as possible.

Slowing $^4$He* atoms from a longitudinal velocity of up to 1050\,m/s ($\sim$30\% of the atoms
in the atomic beam) we trap typically $2\times10^9$ atoms at a 
temperature of 1\,mK.
Alternatively, using a pure $^3$He* source, we trap up to $1\times10^9$ $^3$He* atoms
in the $F=3/2$ state at the same temperature. The number of trapped $^3$He* atoms is
quite large considering that only a few percent of the $^3$He* atoms can be slowed due to 
the larger average velocity of $^3$He* atoms leaving the source; we, however, increased the
loading time and used larger laser power (compared to loading the $^4$He* MOT).

To trap a mixture of $^4$He* and $^3$He* atoms we use two lasers. Our $^4$He laser is a
250\,mW LNA laser while our $^3$He laser is a 1\,W Ytterbium fiber laser. The output of both
lasers is split, sent through various acousto-optic modulators (AOM's) and
overlapped on several beam splitters to provide the collimation/deflection
beam at zero detuning, the Zeeman slower beam at -250\,MHz detuning and the MOT beams
at -40\,MHz detuning. A complication in magneto-optical trapping is
the optical pumping that occurs in $^3$He* due to the $^4$He laser beams~\cite{Stas04}. We found
that optimal trapping is obtained using a repumper beam close to the C$_2$ transition,
simply produced by double passing our Zeeman slower AOM.
We trap a mixture of up to $7\times10^8$ $^3$He* and $1.5\times10^9$ $^4$He* atoms simultaneously
at a temperature of $\approx$1\,mK with this setup~\cite{Tychkov06}.

\subsection{Homonuclear and heteronuclear collisions in a MOT}   
From studies of the ion production in combination with absorption imaging it is possible
to extract two-body loss rate constants in an unpolarized cloud of He* atoms at
temperatures of $\sim$1\,mK. Two different loss rates may be studied.
In the presence of MOT light, losses are dominated by light-assisted collisions, while
in the dark all atoms are in the metastable ground state and loss rates are much smaller.
Experimentally these losses are studied using an MCP detector that attracts all ions
produced by PI/AI, a second MCP detector that measures the temperature, and absorption imaging to determine 
the number of metastables
in the MOT as well as the density profile of the 
cloud. The loss rate for collisions in the dark is determined by
measuring the reduction in the ion signal when the MOT light is switched off for a very short
time, short enough not to disturb the density profile in the MOT.

The goal of the research, which was performed in a setup similar to the setup used for 
magnetic trapping en cooling towards degeneracy, 
was to investigate the difference in PI/AI loss rate constant for 
homonuclear bosonic $^4$He*-$^4$He* collisions, fermionic $^3$He*-$^3$He* collisions and
heteronuclear $^4$He*-$^3$He* collisions, due to quantum statistics. 
For collisions in the presence of MOT light the
difference between the homonuclear loss rate constant for $^4$He* and $^3$He* turned out
to be small as many partial waves contribute to the loss rate constant. For our typical
MOT parameters the two-body loss rate coefficient were 
$3.3(7)\times10^{-9}$\,cm$^3$/s
for $^4$He* and $5.5(8)\times10^{-9}$\,cm$^3$/s for $^3$He*~\cite{Stas06}. These numbers are not
directly comparable to other experimental values as they strongly depend on the MOT laser intensity
and detuning. 

Also in the case of collisions in the dark we did not find large differences in
the Penning ionization rate despite the fact that quantum statistics should
play a prominent role now as only the lowest partial waves contribute.
However, in the dark (when all atoms have decayed
to the metastable state), the atoms are distributed over all magnetic substates
and the difference between the $^4$He* and $^3$He* loss rate constant is 
not very large. We both measured (at the temperature of the MOT), and calculated
(for a large range of temperatures) these loss rate constants. In the
calculations we included quantum threshold behavior, applied Wigner's
spin-conservation rule and incorporated the required quantum statistical
symmetry requirements in a single channel  model~\cite{Stas06, McNamara06b}.
\begin{figure}[t]
\begin{center}
\scalebox{1.0}{\includegraphics{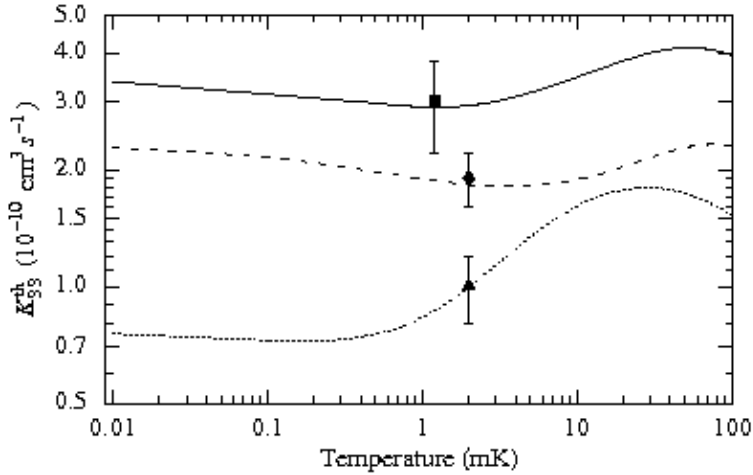}}     % includes figure foo.eps
\end{center}
\caption{Theoretical loss rate coefficients (in the dark) as a function of temperature and
measured values; $^3$He-$^3$He: dashed curve and experimental point (diamond),
$^4$He-$^4$He: dotted curve and experimental point (triangle), $^3$He-$^4$He:
solid curve and experimental point (square).
\label{PI:combined}}
\end{figure}
In Fig.~\ref{PI:combined} the calculated loss rate constants for both the homonuclear
cases and the heteronuclear case are plotted as a function of temperature.
Our measurements are also included and fall on the theoretical curves.
The error in the measurements in the homonuclear case are primarily due to
the measurement of the density profile of the single-isotope MOT. The error
in the heteronuclear case is (only slightly) larger. This can be understood
from the fact that the heteronuclear loss rate in a two-isotope MOT (TIMOT) 
is measured from a total loss rate
that is a combination of homonuclear and heteronuclear losses.

As a bonus we have deduced the fraction of triplet metastables in our beam leaving
the DC discharge. This fraction turns out to be 0.01\%~\cite{Stas06}. We have deduced this number
by measuring the linear losses from our $^4$He* MOT due to collisions with fast ground state $^4$He
atoms from the source (we did not use a deflection zone in this experiment). 
From the known total cross section
a ground state beam intensity of $4\times10^{18}$\,s$^{-1}$ can be calculated, while the
intensity of the $^4$He* beam is $4\times10^{14}$\,s$^{-1}$ for the discharge conditions of
our source.

\section{Magnetic trapping and one-dimensional Doppler cooling}
\begin{figure}[b]
\begin{center}
\scalebox{0.8}{\includegraphics{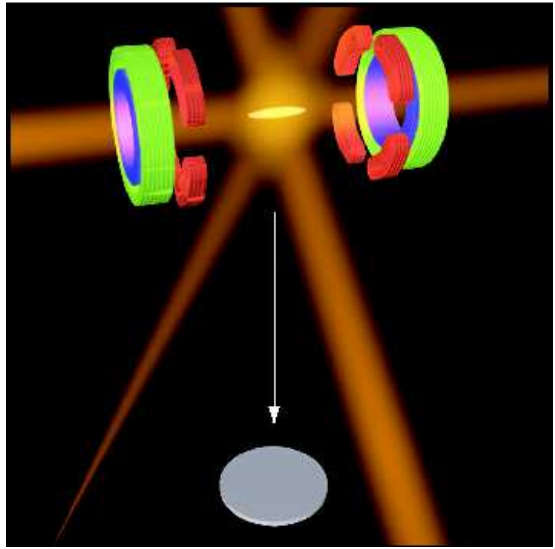}}     % includes figure foo.eps
\end{center}
\caption{Experimental setup for magneto-optical trapping, magnetic trappping and
detection on a microchannel plate detector.
\label{Trap}}
\end{figure}
In order to proceed towards evaporative and sympathetic cooling the atoms
have to be transferred from the MOT to a magnetic trap. Our trap is a cloverleaf trap 
with bias compensation. Its geometry, including the MOT laser beams, 
is shown in Fig.~\ref{Trap}. The coils are
positioned in re-entrant windows that are separated by 3\,cm. The vertical direction is used
for absorption imaging as well as for time-of-flight (TOF) measurement (using an MCP 
positioned 17\,cm below the trap) of the
velocity distribution and atom number. To measure an absorption image, the MCP, 
which is mounted on a translation stage, is moved out of the
vertical laser beam.

To load the magnetic trap all currents and laser beams that produce the MOT are
switched off. Next a weak axial magnetic field is applied and the atoms are spin-polarized
by a short laser pulse along the field direction. The currents for the cloverleaf trap
and the one-dimensional (1D) laser cooling beam are then switched on simultaneously; $\sim$60\% 
of the $^4$He* atoms are transferred to the magnetic trap and a 
temperature of 0.15\,mK
is reached. 1D Doppler cooling turned out to be very efficient, both for $^4$He* and
$^3$He*~\cite{Tychkov06}. The cooling mechanism in this case relies on standard and fast Doppler cooling along
the axial (magnetic field) direction, simply implemented by retroreflecting a very
weak circularly polarized laser beam, about one linewidth red-detuned from resonance at
the center of the trap. The bias magnetic field was 24\,G to prevent optical pumping to
unwanted magnetic substates. Cooling in the radial direction is much slower and
relies on absorption of the red-detuned fluorescence photons; optimum
results (low temperature and no atoms lost during Doppler cooling)
were obtained for 2\,s cooling. For 1D Doppler cooling
to work, the optical density has to be large; we found that it only works well for
more than $10^8$ atoms in the trap. The resulting
phase-space density of our $^4$He* atoms was $\approx$10$^{-4}$, a factor 600 higher than
without 1D Doppler cooling. The lifetime of the atoms in our trap is fully
determined by collisions with background gas. Typically a lifetime of 2 minutes is 
measured. No indications of two-body losses are found.

To increase the collision rate the mixture (or pure cloud) is adiabatically compressed
by reducing the bias magnetic field to a value $\sim$3\,G. The trap frequencies are
then 273 (237)\,Hz in radial direction and 54 (47)\,Hz in axial direction for $^3$He ($^4$He).
The trap thus has a cigar shape.

In the following we first discuss our results for evaporative cooling and BEC of $^4$He*
as these are crucial to the success of sympathetically cooling $^3$He* to quantum degeneracy,
which will be discussed subsequently.

\section{Bose-Einstein condensation of helium-4}
\begin{figure}
\begin{center}
\vspace{3mm}
\scalebox{0.85}{\includegraphics{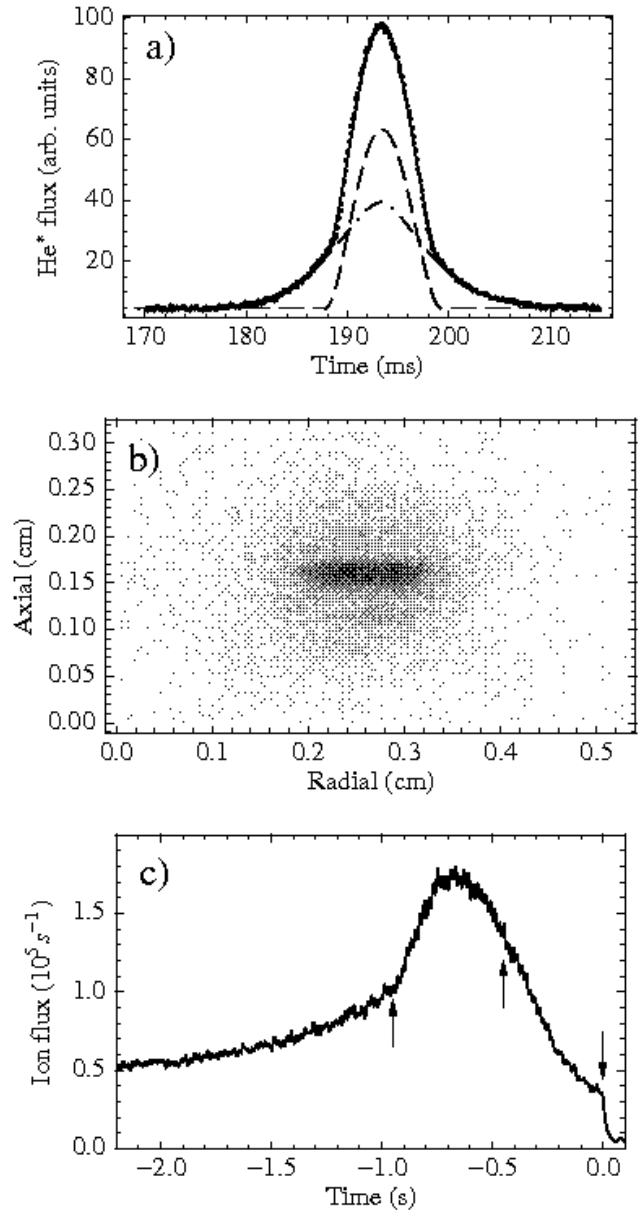}}     % includes figure foo.eps
\end{center}
\caption{Observation of BEC, (a) on the He* microchannel plate detector; the dashed fit
shows the condensed fraction and the dashed-dotted fit the thermal component, (b) on a
CCD camera; after 19~ms expansion a round thermal cloud surrounding an elliptical
condensate is visible, (c) on the ion microchannel plate detector; the condensate
starts to grow at $t=-0.95$~s, at $t=-0.45$~s the RF ramp ends and at $t=0$ the trap is
switched off.
\label{BEC}}
\end{figure}
The ultracold $^4$He* cloud at a phase space density of $\approx$10$^{-4}$ is further cooled
by RF-induced evaporative cooling. For this purpose an RF coil
consisting of two windings (diameter 4\,cm) is positioned 7\,cm from the trap center
just behind the Zeeman slower inside the vacuum. A single exponential ramp, starting
at 50\,MHz and cut at $\approx$8.5\,MHz, cools the atoms in 12\,s to BEC
\cite{Tychkov06}. The temperature
at the onset of BEC is $\approx$2\,$\mu$K. We can also reach BEC in a much shorter ramp,
as short as 2\,s, however at the cost of atom number. The beauty of our experiment
is that we have observed BEC by applying three different
detection techniques \cite{Tychkov06} (see Fig.~\ref{BEC}). The most sensitive
technique is a TOF analysis of the expanding cloud after release from the trap,
measured on the He* MCP mounted below the trap.
A TOF spectrum is shown in Fig.~\ref{BEC}a. It clearly shows the typical double
structure: a broad thermal velocity distribution together with the inverted
parabola of a BEC in the Thomas-Fermi limit. When we move the MCP detector
away in horizontal direction, we measure the absorption image shown in 
Fig.~\ref{BEC}b. This shows the typical signature of BEC in expansion: 
a round thermal cloud and the 
elliptical shape of the condensate. In situ detection of the sudden appearance of a
condensate is possible by observing the ion production in the cloud during evaporative 
cooling. When a BEC forms, the density in the center of the trap suddenly increases,
giving rise to increased two- and three-body PI losses. We see this effect
clearly on the second MCP detector that attracts all ions (Fig.~\ref{BEC}c).

Analysis of the TOF spectra (on CCD and MCP detectors) allows us to measure the
expansion of the condensate in all three spatial dimensions and
determine the number of atoms in the condensate. 
Due, in the main, to detector saturation we can only say that our largest BEC's
contain in-between $1.5\times10^7$ and $4\times10^7$ atoms~\cite{Tychkov06}. These
are large numbers, comparable to some of the largest condensates produced to date,
and at least one order of magnitude larger than other BECs of He*. In BEC decay
experiments~\cite{Tychkov06}, we were able to detect a condensate up 
to 75\,s
after it was formed, due to the wide range and high sensitivity of
the MCP detector. 

\section{Fermi degeneracy of helium-3}
\begin{figure}[b]
\begin{center}
\scalebox{0.4}{\includegraphics{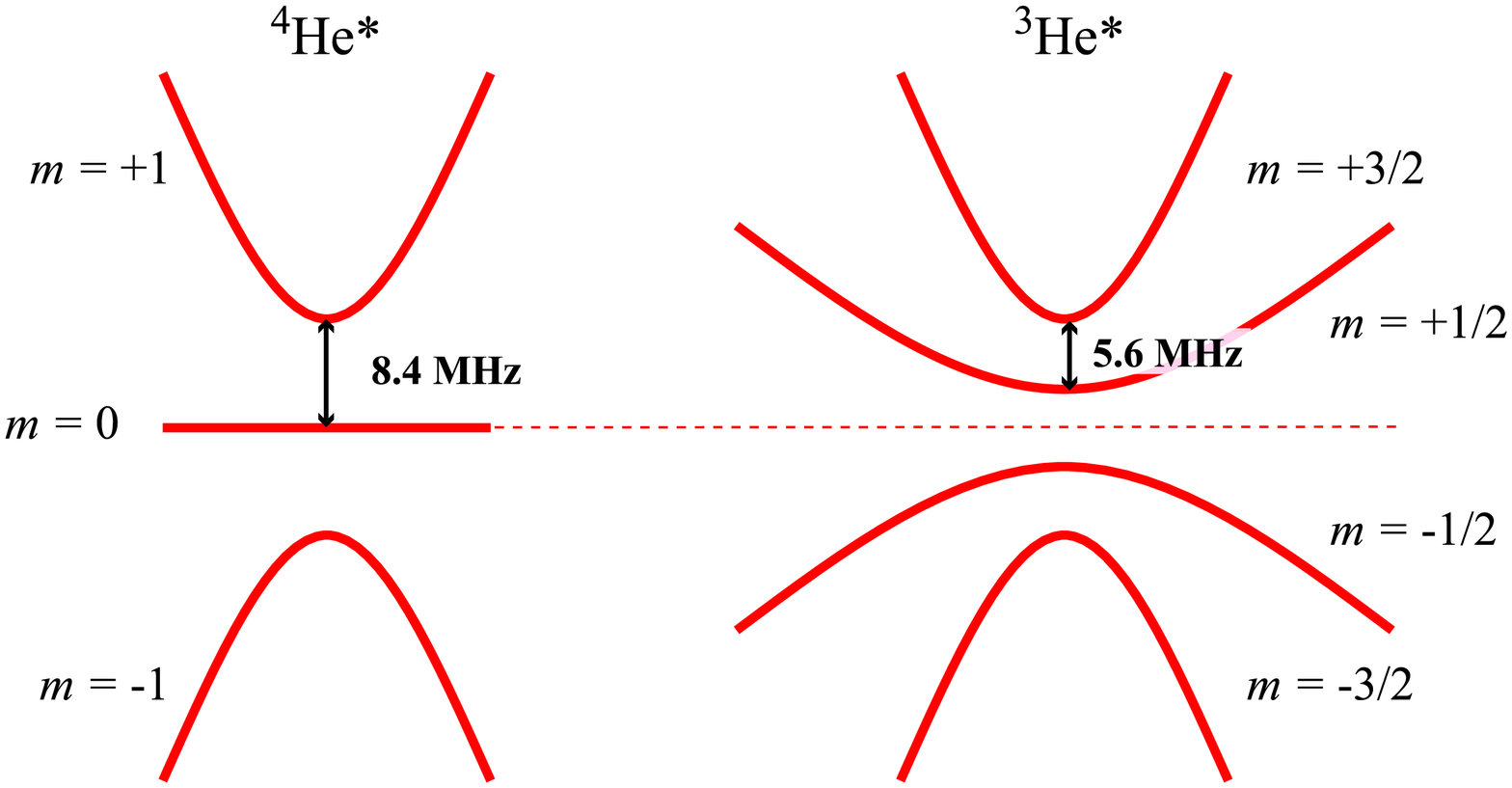}}     % includes figure foo.eps
\end{center}
\caption{Potential energy of the different magnetic substates of both $^4$He* and $^3$He*
in the cloverleaf magnetic trap. In the center of the trap the $m=+1$ $^4$He atoms are
(at $T=0$) removed with RF at a frequency of 8.4~MHz while the $m=+3/2$ $^3$He* atoms can
be removed at a frequency of 5.6~MHz.\label{RFoutput}}
\end{figure}
We discussed in Sect.~4 that we can trap a mixture of up to $7\times10^8$ $^3$He*  
and $1.5\times10^9$ $^4$He* atoms simultaneously at a temperature of 
$\approx$1\,mK in our TIMOT.
Given the size of our $^4$He* reservoir it is impossible to cool so many fermions 
to quantum degeneracy by sympathetic cooling. Our BEC experiments
show that we cross the BEC threshold with typically $\approx5\times10^7$ bosons. We
therefore reduce the number of fermions in our TIMOT to $\approx10^7$ by changing
the composition of the isotopic mixture in our helium reservoir to 90\% $^4$He and 10\% $^3$He or
reducing the loading time of the $^3$He* MOT. The number of $^3$He* atoms should not be
chosen too low as then the 1D Doppler cooling does not work as well as at
higher number due to loss of optical density. There is clearly a compromise here
which still needs optimization. After 1D Doppler cooling and compression we
have a mixture in thermal equilibrium and a collision rate suitable for
efficient evaporative cooling.
We then perform RF-induced evaporative
cooling of $^4$He* and sympathetically cool $^3$He* in the
$|3,2,+3/2\rangle$ state. Measuring the number of $^4$He* atoms and $^3$He* atoms during
evaporative cooling we see the typical reduction in the number of $^4$He* atoms while
the number of $^3$He* atoms stays roughly constant. This
shows that sympathetic cooling works efficiently, as expected from the
large value of the heteronuclear scattering length.
We have not (yet) observed the $^3$He* atoms by absorption
imaging, in stead we rely upon TOF measurements in our determination of
temperature and number of atoms. Of course an MCP does not discriminate between the
two isotopes so in order to follow the sympathetic cooling process we measure the 
number of atoms from one isotope after removing the other from the trap. At
high temperature ($T>$20\,$\mu$K) we use resonant laser light for this purpose. 
At these temperatures the number of $^4$He* atoms is much larger than the number of
$^3$He* atoms so the $^4$He* number of atoms and temperature are not very difficult to
measure. On the other hand, it is very difficult to measure the number of $^3$He* at these
'high' temperatures. To measure a $^3$He* TOF we need to push a much larger number of
$^4$He* away, while we know that $^4$He* light affects the $^3$He* cloud by off-resonant
excitation of the C$_9$ transition (see Fig.~\ref{A1}). This 'pushing' technique therefore
is not very accurate. Luckily,
when the temperature has decreased below $\approx$20\,$\mu$K we can selectively
remove either $^4$He* or $^3$He* by applying a short RF ramp. This is illustrated
in Fig.~\ref{RFoutput}. When we want to measure either $^4$He* or $^3$He* we
generate a short RF sweep and cut through the distribution of the other isotope.
The frequency span of this ramp is, at temperatures below 20\,$\mu$K, small enough not to disturb
the isotope that we want to study. We have noted that $^3$He* is efficiently
removed despite the fact that two RF photons have to be absorbed to remove it 
from the trap. We are thus able to measure the mixture by simply releasing it from
the trap or by measuring each isotope separately. Applying a magnetic field gradient,
it is possible to separate both isotopes in time as well.
\begin{figure}[t]
\begin{center}
\scalebox{0.8}{\includegraphics{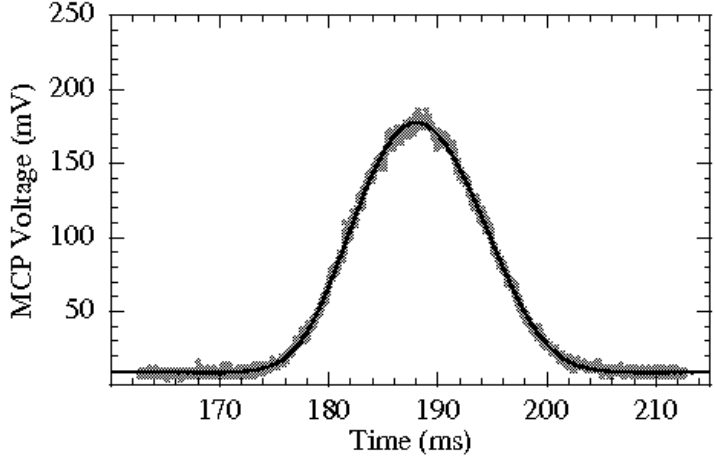}}     % includes figure foo.eps
\end{center}
\caption{Time-of-flight signal of a degenerate Fermi gas of $^3$He* atoms together
with a fit to a Fermi-Dirac velocity distribution. The fit shows that in this case
the number of $^3$He* atoms is $2.1\times10^6$, the temperature $0.8\,\mu$K and the
degeneracy parameter $T/T_F=0.45$.
\label{DFG}}
\end{figure}

Our first experiments aimed at producing a degenerate Fermi gas (DFG), so we applied a
single RF ramp with end frequency below 8.4\,MHz. This cools and effectively removes all $^4$He*
atoms from the trap while the $^3$He* is sympathetically cooled and remains in the trap.
A 'typical' TOF spectrum obtained in this way is shown in Fig.~\ref{DFG}.
A perfect fit is obtained when we fit this spectrum with a Fermi-Dirac TOF function;
a Maxwell-Boltzmann classical TOF function clearly does a less good job.
This TOF corresponds to a degenerate Fermi gas with a temperature of 
0.8\,$\mu$K and
degeneracy parameter $T/T_F=0.45$, with N$_3=~2.1\times10^6$ $^3$He* atoms in
the $|3/2,+3/2\rangle$ state. To calculate the Fermi temperature we used the standard
formula:
\begin{equation}
k_BT_F=h(6N_3\nu_r^2\nu_a)^{1/3}
\end{equation} 
with $k_B$ Boltzmann's constant, $h$ Planck's constant and $\nu_a$ and $\nu_r$ the axial
and radial trap frequencies.
\begin{figure}[b]
\begin{center}
\scalebox{0.8}{\includegraphics{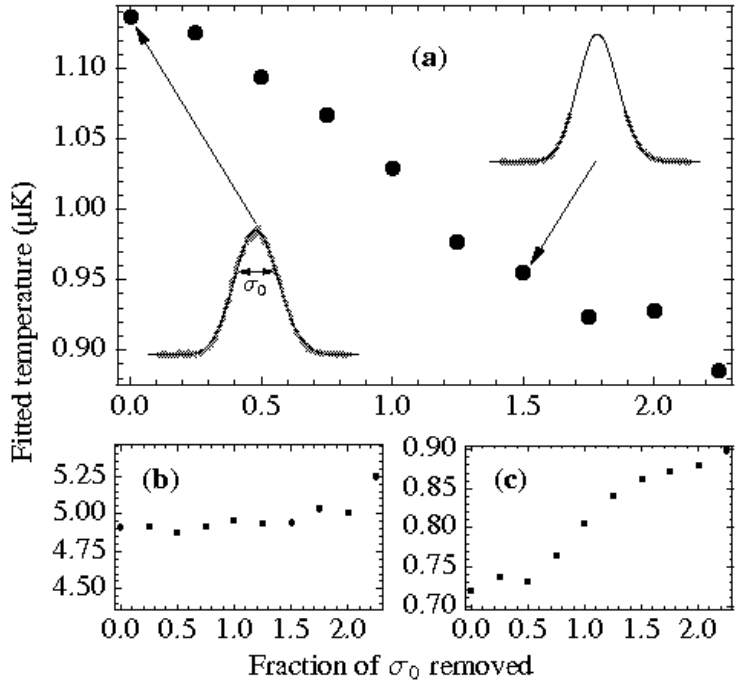}}     % includes figure foo.eps
%\scalebox{0.8}{\includegraphics[viewport=0 55 350 367]{Schreckfigure.eps}}     % includes figure foo.eps
\end{center}
\caption{Temperatures obtained by repeatedly fitting, with a classical Maxwell-Boltzmann
distribution, three time-of-flight spectra from which increasingly large central fractions have been removed:
(a) a degenerate $^3$He* Fermi gas at $T/T_F=0.5$, (b) a cloud of $^4$He* with $T>>T_C$, and
(c) a cloud of $^4$He* atoms just above $T_C$.\label{Schreckfigure}}
\end{figure}
In alkali absorption spectra on a CCD camera the TOF distribution in two dimensions
is observed and the signal is integrated over the third (line-of-sight) dimension.
This already makes it difficult to see the difference between a Maxwell-Boltzmann
distribution and a Fermi-Dirac distribution. In our case we effectively integrate
over two dimensions (the surface of the MCP detector) 
and only the expansion in the radial direction is measured.
This makes the small difference between these two velocity distributions even
smaller. However, our TOF spectra show a very good signal-to-noise ratio and we can
fit far into the wings of the distribution. 
To show that indeed the Fermi-Dirac velocity distribution is best
in fitting a degenerate Fermi gas and to compare these results also with similar
results for bosons close to BEC we measured four TOF spectra, two (for $^3$He* and $^4$He*)
close to degeneracy and two (for $^3$He* and $^4$He*) far above. We fitted these four
TOF's with a gaussian (for a Maxwell-Boltzmann distribution) and extracted the temperature
and rms width $\sigma_0$. 
Then, as described by Schreck \cite{Schreck02}, we removed
increasingly larger central fractions of the TOF spectrum and fitted it again. The temperature
we extract as a function of the fraction of $\sigma_0$ removed is plotted in
Fig.~\ref{Schreckfigure}, in Fig.~\ref{Schreckfigure}a for $^3$He* below $T_F$,
in Fig.~\ref{Schreckfigure}b for $^4$He* far above $T_C$ (a similar spectrum is
obtained for $^3$He* above $T_F$) and in Fig.~\ref{Schreckfigure}c for $^4$He*
just above $T_C$. For a cloud displaying Fermi-Dirac statistics the population
of low energy states is suppressed due to the Pauli exclusion principle and
fitting a gaussian to the whole TOF will lead to an overestimation of the
temperature. By fitting only the wings this overestimation is reduced and the
extracted temperature will fall as a function of the fraction of $\sigma_0$ 
removed in the fit. Fig.~\ref{Schreckfigure}a clearly shows this. 
The opposite behaviour was observed
for the bosons where the low-energy states show enhanced population due to Bose enhancement
(Fig.~\ref{Schreckfigure}c). A thermal cloud should see no dependence on the fraction of 
$\sigma_0$ removed and this
is shown clearly in Fig.~\ref{Schreckfigure}b, which was recorded at a temperature
of $\sim$5\,$\mu$K, far above $T_C$ (and $T_F$). From Fig.~\ref{Schreckfigure}a we again recover
$T/T_F$=0.5 when we fit only the wings removing the central part up to 
1.75\,$\sigma_0$.
The large value for the theoretical heteronuclear scattering length of 
+29\,nm suggests
that thermalization should be very efficient. Indeed, we can produce a degenerate
Fermi gas with an RF ramp of 2.5\,s, albeit with a slightly larger degeneracy
parameter. For such a ramp we measure $T/T_F=0.75$, with $N_3=4\times 10^6$. In
this way we thus produce larger clouds of degenerate $^3$He* gas.

\begin{figure}[b]
\begin{center}
\scalebox{0.8}{\includegraphics{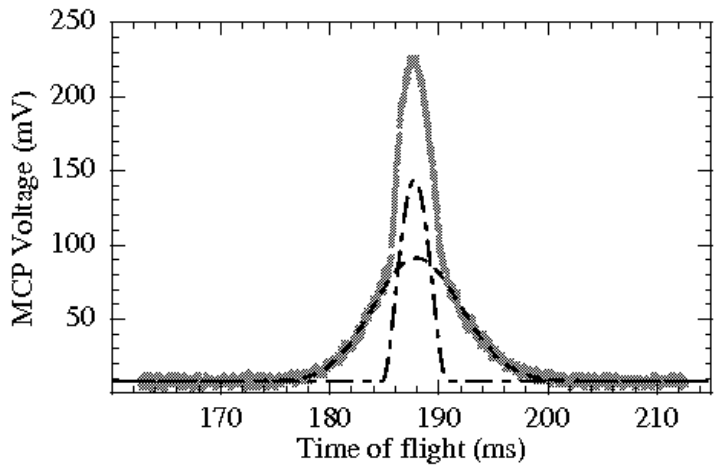}}     % includes figure foo.eps
\end{center}
\caption{Time-of-flight spectrum for a degenerate mixture of $4.2\times10^5$ ($T/T_F=0.5$ $^3$He*
and $1\times10^5$ $^4$He* atoms. The dashed curve is a Fermi-Dirac velocity distribution fitted
to the wings while the dashed-dotted curve is a fit of a pure Bose-Einstein condensate to
the central part.\label{Degenerate_mixture}}
\end{figure}
\section{Mixtures}
It is straightforward to continue along the lines described in the previous section
to realize a quantum degenerate mixture of $^4$He* in the 
$|1,+1\rangle$ state and
$^3$He* in the $|3/2,+3/2\rangle$ state. The only difference is the end frequency of the
RF ramp which now is to be chosen slightly above the bottom of the 
$^4$He* trap, i.e.,
slightly above 8.4\,MHz. This results in a pure BEC, sitting (presumably) in the center of
a degenerate Fermi gas. An observed TOF for this situation is shown in
Fig.~\ref{Degenerate_mixture}. This specific mixture contained $N_3=4.2\times 10^5$
$^3$He* fermions and a condensate with $N_4=1\times 10^5$ $^4$He* atoms
at a temperature corresponding to $T/T_F=0.5$. The two theoretical curves
represent the Fermi-Dirac velocity distribution, fitted to the wings, and the
inverted parabola of a pure BEC in the Thomas-Fermi limit.
These experiments are still in a preliminary phase but we may already conclude from
our observations that the lifetime of a condensate ($\tau_C$) in the presence of a DFG of $^3$He*
is significantly shorter than in case of a pure condensate. We find 
$\tau_C=1$\,s for a pure
condensate~\cite{Tychkov06} and $\tau_C\sim 10$\,ms for the mixture. This can be understood
when indeed, as suggested in Sect.~3, the three-body loss rate $K_3^{BBF}$ is
large.

\section{Prospects}
\begin{figure}[b]
\begin{center}
\scalebox{0.6}{\includegraphics{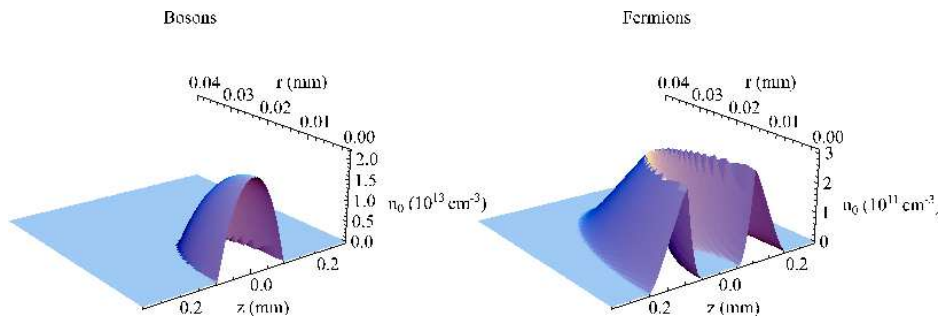}}     % includes figure foo.eps
\end{center}
\caption{$T=0$ calculation of the spatial distribution of a Bose-Einstein condensate of $10^5$ $^4$He*
atoms in the $|1,+1\rangle$ state (left) and a degenerate Fermi gas of $10^6$ $^3$He* atoms
in the $|3/2,+3/2\rangle$ state (right) in our cloverleaf magnetic trap. The strong repulsion due
to the large and positive 3-4 scattering length causes the fermions to be repelled out of the
core of bosons.\label{separation}}
\end{figure}

Now that we have seen that we can produce a degenerate Fermi gas of metastable $^3$He atoms as well as
a degenerate boson-fermion mixture of metastable atoms, it is interesting to discuss
what these gases may be good for. We have a system with both a naturally large
and positive scattering length and this offers some interesting possibilities.
It may be possible to observe phase separation~\cite{Molmer98} in this system. For the large scattering
lengths of this system our mixture may already exhibit phase separation, although
we have not demonstrated this. A $T=0$ calculation based on the Gross-Pitaevskii equation
for the bosons and the Thomas-Fermi equation for the fermions \cite{Marten05}, shown in Fig.~\ref{separation},
shows that the fermions are expected to be expelled from the hard core of bosons already
at the densities calculated for observed particle numbers at degeneracy. 
As discussed in Sect.~3 it will be very difficult to make spin mixtures of ultracold $^3$He* atoms.
This hampers observation of BCS-like phenomena observed for the Li and K systems.
However, as the 3-4 scattering is so large, we need only a very small tuning of the scattering
lengths to open up the p-wave window for Cooper pairing of 
$|3/2,+3/2\rangle$ $^3$He* atoms.
This pairing is mediated by density fluctuations in the $^4$He* cloud in a boson-fermion
mixture~\cite{Efremov02}. It seems not entirely impossible to use an optical Feshbach resonance for this
purpose. Also loading such a mixture in an optical lattice offers many interesting
new possibilities~\cite{Lewenstein04}. First this offers a new system for the observation of exotic
new phases and secondly a boson-fermion mixture in an optical lattice will be a
naturally clean system with at most one boson per lattice site due to the fast
Penning ionization that will occur when two $^4$He* atoms or a $^4$He* and a $^3$He* atom
occupy the same lattice site.

Another important application of ultracold $^4$He* and $^3$He* will be in metrology.
A source of ultracold metastable helium atoms provides an almost ideal opportunity to
measure with high spectral resolution narrow transitions from the 
$2\,^3\rm{S}_1$ state.
Of particular interest are the electric-dipole allowed 1083\,nm line 
(to $2\,^3\rm{P}$) 
with a natural linewidth of 1.6\,MHz and the magnetic-dipole allowed 
1557\,nm line 
(to $2\,^1\rm{S}$), with a natural linewidth of 8\,Hz \cite{Baklanov97, Leeuwen06}. Observation of these transitions
in a 1D optical lattice will allow virtually Doppler-free excitation offering a
resolution below 1 kHz. This allows tests of bound state QED in two-electron
systems, and a measurement of differences in nuclear charge radii as well as nuclear masses.
The latter will be possible if we measure the transition isotope shift 
of the 1557\,nm line, which at present can be
calculated from theory with an accuracy better than 1\,kHz
\cite{Leeuwen06}.

Perhaps the most promising experiment that can be performed with an ultracold
cloud of $^3$He* atoms is the analogue of the recently realized Hanbury-Brown and
Twiss experiment for $^4$He* atoms in Orsay \cite{Schellekens05}. For $^4$He*
bunching was observed in an ultracold gas of bosons, close to BEC, while for
a BEC this bunching was absent, just as expected for a coherent source of atoms.
This in complete analogy to the classical light source and the laser. We have the
perfect source to repeat this experiment with fermions and 
to observe anti-bunching for an ultracold gas of $^3$He*. Actually, on July 14,
two weeks after the end of the School, we indeed observed antibunching in our setup,
in a collaboration with the Orsay group, who brought their position-sensitive MCP
detector to Amsterdam~\cite{Westbrook06}.

All these proposed experiments will profit from the possibility to observe an ultracold
cloud of metastable atoms using the three detection techniques: absorption
imaging (so far only used on the bosons), TOF measurements using an MCP and
ion detection (also using an MCP detector). The last detection technique was so far only used for the bosons as well.

\end{document}